\documentclass[12pt]{article}

\input{epsf}
\usepackage{epsfig}
\usepackage{amssymb}
\usepackage{amsfonts}
\usepackage{amsbsy}

\newcommand\atmp[3]
        {{\it Adv.\ Theor.\ Math.\ Phys.\ }{\bf #1} (#2) #3}

\newcommand\jmp[3]
        {{\it J.\ Math.\ Phys.\ }{\bf #1} (#2) #3}

\newcommand\cqg[3]
        {{\it Class.\ and Quant.\ Grav.\ }{\bf #1} (#2) #3}

\newcommand\jhep[3]
        {{\it J. High Energy Phys.\ }{\bf #1} (#2) #3}

\newcommand\npb[3]
        {{\it Nucl.\ Phys.\ }{\bf B #1} (#2) #3}

\newcommand\plb[3]
        {{\it Phys.\ Lett.\ }{\bf B #1} (#2) #3}

\newcommand\prd[3]
        {{\it Phys.\ Rev.\ }{\bf D #1} (#2) #3}

\newcommand{\hepth}[1]{{\tt hep-th/#1}}

\def\IC{\mathbb{C}}

\def\IZ{\mathbb{Z}}
\def\IR{\mathbb{R}}

\def\s*{\boldsymbol{*}}

\def\sx{{\bf x}}

\def\sJ{{\bf J}}

\def\CM {{\cal M}}
\def\CN {{\cal N}}

\def\CF {{\cal F}}

\def\CA{{\cal A}}
\def\CK{{\cal K}}

\newcommand{\sect}[1]{\setcounter{equation}{0}\section{#1}}

\renewcommand{\Im}{{\rm Im }}
\renewcommand{\Re}{{\rm Re }}

\def\one{{\hbox{ 1\kern-.8mm l}}}

\def\bz{{\bar z}}

\newcommand{\N}{{\cal N}}

\def\bb{\bar{b}}
\def\bz{\bar{z}}

\def\bOmega{\bar{\Omega}}
\def\bpartial{\bar{\partial}}

\def\dop{d}

\begin{document}

\begin{titlepage}
\begin{flushright} RUNHETC-2003-10
\end{flushright}

\vfill
\begin{center}
{\LARGE\bf Exact solutions for supersymmetric stationary black hole
composites}
\\
\vskip 20mm {\bf Brandon Bates} ${}^1$ and {\bf Frederik Denef} ${}^2$ \\
\vskip 7mm ${}^1$ Department of Physics, Columbia University \\
New York, NY 10027, USA \\
{\tt bdbates@phys.columbia.edu} \\
\vskip 5mm
${}^2$ Department of Physics and Astronomy, Rutgers University \\
Piscataway, NJ 08855-8019, USA \\
{\tt denef@physics.rutgers.edu}
\end{center}
\vfill

\begin{quote}

\begin{center}
{\bf Abstract}
\end{center}
\vskip 5mm

Four dimensional ${\cal N}=2$ supergravity has regular,
stationary, asymptotically flat BPS solutions with intrinsic
angular momentum, describing bound states of separate extremal
black holes with mutually nonlocal charges. Though the existence
and some properties of these solutions were established some time
ago, fully explicit analytic solutions were lacking thus far. In
this note, we fill this gap. We show in general that explicit
solutions can be constructed whenever an explicit formula is known
in the theory at hand for the Bekenstein-Hawking entropy of a
single black hole as a function of its charges, and illustrate
this with some simple examples. We also give an example of
moduli-dependent black hole entropy.

\end{quote}

\vfill


\end{titlepage}

\sect{Introduction}

The analysis of BPS states in various compactifications of string
theory has been of fundamental importance in exploring
non-perturbative phenomena, dualities and quantum geometry. In
particular in type II theory compactified on a Calabi-Yau
threefold, where BPS states have a description as D-branes wrapped
on various supersymmetric cycles (and generalizations thereof),
this study has revealed some remarkable physical and mathematical
structures. The low energy effective theory of such a Calabi-Yau
compactification is a four-dimensional $\CN=2$ supergravity
theory, coupled to a number of massless abelian vector- and
hypermultiplets, and in this theory BPS states have a description
complementary to the D-brane picture as solutions to the field
equations preserving $\CN=1$ supersymmetry. The simplest solutions
of this kind are spherically symmetric black holes, first studied
in \cite{FKS}. As noted in \cite{M}, not all charges support such
solutions. This is natural, since also in the full string theory,
the BPS spectrum is only a subset of the full charge lattice.
However, it turns out that the true BPS spectrum and the spectrum
of spherically symmetric black holes do not match
\cite{D,branessugra}, the latter being too small. To reconcile the
two, one has to drop the restricion to spherically symmetric
solutions with a single charge center, and consider multicentered
composites as well \cite{branessugra}. Indeed, $\CN=2$
supergravity has regular BPS ``bound state'' solutions describing configurations
of distinct (typically mutually nonlocal) charges at rest at
certain equilibrium separations from each other. These solutions
are in general stationary but non-static, as they can carry an
intrinsic angular momentum, much like the monopole-electron system
in ordinary Maxwell electrodynamics. Furthermore their existence
is subject to certain moduli-dependent stability conditions,
matching similar conditions appearing in the D-brane description
of BPS states \cite{DFR,DFR2,Dcat}. With these ingredients,
supergravity predictions for the BPS spectrum
of the full string theory were given in \cite{DGR}.

The equations of motion for general stationary BPS configurations
were derived in \cite{BLS,branessugra,CWKM}. In
\cite{branessugra}, non-static black hole composites were
considered as solutions, and some of their properties were
analyzed directly from the equations, assuming existence. A proof
of existence was later given in \cite{montreal}, however without
giving explicit analytic expressions for the metric, scalars and
vectors. In what follows, we will show how such closed form
expressions can be found. Perhaps somewhat surprisingly, it turns
out that the full solution can be built from a single function,
namely the Bekenstein-Hawking entropy (i.e.\ the horizon area) of
a single BPS black hole as a function of its charges. If the
latter is known analytically, the same is true for the complete
space-time geometry and all fields involved, for arbitrary values
of the moduli at spatial infinity.

Some examples of non-static multicentered solutions were studied earlier in \cite{CWKM2},
for supergravity with $R^2$-corrections, focusing mainly on how to take the curvature
corrrections into account in an iterative approximation scheme. The zeroth order part
of those results can be obtained as a special case of the general construction outlined
in this note. An explicit expression for the off-diagonal part of the metric was not given in
\cite{CWKM2}. We show how this part can be obtained in closed form without too much additional
effort, and note that requiring its regularity leads to constraints on the positions of the
centers and to certain stability conditions, confirming general expectations \cite{branessugra}.

\sect{Notation and setup}

\subsection{General formalism}

We will assume that the supergravity theory under consideration
arises from compactification of type IIB string theory on a
Calabi-Yau threefold, thus giving a concrete
geometric interpretations to the various quantities involved. The
generalization to arbitrary supergravity theories determined by
abstract special geometry data will be obvious.

Compactification of IIB on a Calabi-Yau manifold $X$ gives as
four-dimensional low energy theory $\N = 2$ supergravity coupled
to $n_v = h^{1,2}(X)$ massless abelian vectormultiplets and $n_h =
h^{1,1}(X)+ 1$ massless hypermultiplets, where the $h^{i,j}(X)$
are the Hodge numbers of $X$. The hypermultiplet fields will play
no role in the following and are set to arbitrary constant values.

The complex scalars in the vector multiplets are the complex
structure moduli of $X$. The geometry of the corresponding scalar
moduli space $\CM$, parametrized with $n_v$ coordinates $z^a$, is
special K\"ahler~\cite{SG}. In what follows we recall some general
facts and useful formulas in special geometry (in the IIB
geometric setting).

The basic objects in special geometry are $(i)$ a vector space $V$, here
identified with the $2n_v+2$ dimensional vector space of harmonic
3-forms $H^3(X,\IC)$, for which we pick an arbitrary basis $\{ \Theta_A \}$;
$(ii)$ an antisymmetric bilinear form $\langle
\cdot , \cdot \rangle$, here identified with the intersection
product on $H^3(X,\IC)$,
\begin{equation} \label{intproddef}
 \langle Q_1,Q_2 \rangle = \int_X Q_1 \wedge
Q_2
= Q_1^A \, I_{AB} \, Q_2^B , \quad \mbox{where } I_{AB} = \langle \Theta_A,\Theta_B \rangle;
\end{equation}
and $(iii)$ a $V$-valued holomorphic function\footnote{single-valued only
on the covering space $\widetilde{\CM}$ of $\CM$},
here identified with the holomorphic 3-form on $X$,
\begin{equation}
 \Omega_0(z) = \Omega_0^A(z) \, \Theta_A = I^{AB} X_B(z) \, \Theta_A, \quad \mbox{where } X_A
 = \langle \Theta_A, \Omega_0 \rangle = \int_{\widetilde{\Theta}_A}
\Omega_0,
\end{equation}
with $I^{AB} \equiv (I^{-1})^{AB}$ and $\widetilde{\Theta}_A$ the
3-cycle Poincar\'{e} dual to $\Theta_A$. The vector $X_A$ is
called the holomorphic period vector.

The special K\"ahler metric $g_{a\bb} =\partial_a \bpartial_{\bb} \CK$ on
$\CM$ is derived from the K\"ahler potential
\begin{equation}
\CK(z,\bz) = - \ln i \langle \Omega_0, \bOmega_0 \rangle = - \ln
\left( i \, \bar{X}_A I^{AB} X_B \right).
\end{equation}
It is useful to introduce also the {\em normalized} 3-form and
period vector
\begin{equation}
 \Omega = e^{\CK/2} \Omega_0, \qquad Z_A=e^{\CK/2} X_A
\end{equation}
Note that $Z_A$ has non-holomorphic dependence on the moduli
through $\CK$. The K\"ahler covariant derivative $D_a$ is defined
on these normalized objects as $D_a \equiv \partial_a +
\partial_a\CK/2$. Then $D_a \Omega \in H^{2,1}(X)$,
and since $\Omega \in H^{3,0}(X)$, one has $\langle \Omega,D_a \Omega \rangle = 0$,
$\langle \bOmega,D_a \Omega \rangle = 0$. Furthermore
\begin{equation}
 \langle \Omega,\bar{\Omega} \rangle = -i, \qquad \langle D_a
\Omega,\bar{D}_{\bb} \bOmega \rangle = i \, g_{a \bb}.
\end{equation}

The low energy dynamics of the vector fields is also determined by
special geometry. The type IIB self-dual five-form field strength
$\CF = d \CA$ descends to the four dimensional electromagnetic field
strengths $F^I = d A^I$ and
their magnetic duals $G_I$ by the decomposition
$\CF = F^I \otimes \beta_I - G_I \otimes \alpha^I$, where $\{
\alpha^I,\beta_I \}_{I=1,\ldots,n_v+1}$ is a fixed standard $\langle
\cdot,\cdot \rangle$-symplectic basis\footnote{i.e. $\langle \alpha^I,\beta_J \rangle = \delta^I_J$
and $\langle \alpha_I,\alpha_J \rangle = \langle \beta^I,\beta^J \rangle = 0$.} of harmonic 3-forms on $X$.
The fields $G_I$ and $F^I$
are not independent: they are related by the self-duality
constraint on $\CF$.In the four dimensional context, we refer to the $H^3(X,\IR)$-valued field
$\CF$ as the total electromagnetic field strength.

The lattice of electric and magnetic charges is identified with
$H^3(X,\IZ)$. The origin of
a charge $\Gamma \in H^3(X,\IZ)$ in type IIB string theory (at $g_s \to 0$) is a
D3-brane wrapped around the cycle Poincar\'e dual to $\Gamma$.

\subsection{Example: diagonal $T^6$} \label{sec:diagT6}

Let $X_\tau$ be the diagonal $T^6$ \cite{M} with modulus $\tau$,
that is, $X_\tau=E_\tau \times E_\tau \times E_\tau$, where
$E_\tau$ is the 2-torus with standard complex structure parameter
$\tau=b + i a$ (valued in the upper half plane). This gives a
consistent truncation of the full $\mbox{IIB}/T^6$ theory, if
moreover we only consider charges $\Gamma \in H^3(X,\IZ)$
invariant under the permutation symmetry of the three 2-tori.

Type IIB string theory on $X_\tau$ is mirror (or
T-dual\footnote{By T-dualizing along the horizontal direction in
each $T^2$}) to IIA on $Y=E'_\tau \times E'_\tau \times E'_\tau$,
with $E'_\tau$ the 2-torus with area $a=\Im \, \tau$ and $B$-field
flux $b=\Re \, \tau$ (which together determine the complexified
K\"ahler class of $Y$). There are four charges $\Gamma \in
H^3(X,\IZ)$ invariant under the permutation symmetry, mirror to
D0-, D2-, D4- and D6-branes on the IIA side. Denoting the standard
complex coordinate in the $i$-th $T^2$ by $z^i=u^i+\tau v^i$,
these charges are explicitly:
\begin{eqnarray}
 D0 & \leftrightarrow & - dv^1 \wedge dv^2 \wedge dv^3\\
 D2 & \leftrightarrow & dv^1 \wedge dv^2 \wedge du^3
  + dv^1 \wedge du^2 \wedge dv^3 + du^1 \wedge
  dv^2 \wedge dv^3 \\
 D4 & \leftrightarrow & -du^1 \wedge du^2 \wedge dv^3
  - du^1 \wedge dv^2 \wedge du^3 - dv^1 \wedge
  du^2 \wedge du^3 \\
 D6 & \leftrightarrow & du^1 \wedge du^2 \wedge du^3.
\end{eqnarray}
The holomorphic 3-form on $X$ is $\Omega=dz^1 \wedge dz^2 \wedge
dz^3$. With respect to the (D0,D2,D4,D6)-basis, the period vector
is $X=(1,3\tau,-3\tau^2,-\tau^3)$. So the D0-brane is mirror to a
D3-brane wrapped in the $u^i$-directions, and so on. The
intersection matrix is
\begin{equation}
 I = \left( \begin {array}{cccc} 0&0&0&1\\ 0&0&-3&0\\0&3&0&0
\\-1&0&0&0\end {array} \right),
\end{equation}
the special K\"ahler potential on moduli space is $\CK=-\ln(8 a^3)$,
and the corresponding metric is $g_{\tau\bar{\tau}} = 3/4a^2$.

\sect{BPS equations of motion} \label{sec:bpseom}

In this section we recall the BPS field equations for a general
stationary black hole composite. The $\CN=2$
supergravity plus vector multiplet action is, in units with $G_N=1$,
\begin{equation}
 S_{4d}=\frac{1}{16 \pi} \int d^4 x \sqrt{-h} R \,
  - \, 2 g_{a\bb} \, \dop z^a \wedge *\dop \bz^{\bb} \,\,
  - F^I \wedge G_I \, . \label{S4D}
\end{equation}
The action for a probe BPS particle of charge $\Gamma$ is
\begin{equation} \label{Sp}
 S_{p} = - \int |Z| \, ds + \frac{1}{2} \int \langle \Gamma,\CA \rangle,
\end{equation}
where $Z=Z_\Gamma=\Gamma^A Z_A$.
providing a source for the fields in (\ref{S4D}).
A BPS metric is of the form
\begin{equation} \label{stationarymetric}
 ds^2 = e^{2U} \left(dt + \omega_i dx^i\right)^2 - e^{-2 U} dx^i
 dx^i\,,
\end{equation}
where $U$ and $\omega$, together with the moduli fields $z^a$, are
time-independent solutions of the following equations \cite{branessugra,CWKM}:
\begin{eqnarray}
 2 e^{-U} \Im \left(e^{-i\alpha} \Omega\right) &=& - H \,, \label{mc1}\\
 \s* d \omega &=& \langle d H,H \rangle \,, \label{mc2}
\end{eqnarray}
with $\alpha$ an unknown real function, $H(\sx)$ a given
$H^3(X,\IR)$-valued harmonic function (on flat coordinate space
$\IR^3$), and $\s*$ the Hodge star operator on \emph{flat} $\IR^3$. For
$N$ charges $\Gamma_s$ located at coordinates $\sx_s$,
$p=1,\ldots,N$, in asymptotically flat space, one has:
\begin{equation}
  H = \sum_{s=1}^N \frac{\Gamma_s}{|\sx-\sx_s|}  \, \, - \,  2 \Im\left(e^{-i
  \alpha} \Omega\right)_{r=\infty} \,. \label{Haha}
\end{equation}
The boundary condition on $\alpha$ at $r=\infty$ is that it equals
the phase of the total central charge, $\alpha = \arg Z_\Gamma$, with
$Z_\Gamma = \Gamma^A Z_A$ and
$\Gamma=\sum_s \Gamma_s$.

The total electromagnetic field $\CF=d \CA$ is
furthermore given by
\begin{equation}
 \CA = 2 e^U \Re \left(e^{-i \alpha} \Omega\right) (dt + \omega) + \CA_d,
  \label{bpsmc3}
\end{equation}
where $\CA_d$ is a Dirac magnetic monopole type vector potential obtained from
\begin{equation}
 d \CA_d = - 2 \s* d \left( e^{-U} \Im \left(e^{-i\alpha} \Omega \right) \right)
= \s* d H
\end{equation}

Not all positions of the charges are allowed \cite{branessugra}, as equation (\ref{mc2})
leads to an integrability condition, obtained by acting with $d
\s*$ on both sides of the equation: for all
$s=1,\dots,N$
\begin{equation} \label{distconstr}
\sum_{t=1}^N \frac{\langle \Gamma_s,\Gamma_t
\rangle}{|\sx_s-\sx_t|} = 2 \, \Im\left(e^{-i \alpha} Z_s
\right)_{r=\infty}.
\end{equation}
In the case of just two charges $\Gamma_1$ and $\Gamma_2$, this
simplifies to
\begin{equation} \label{eqsep}
|\sx_1-\sx_2| = \frac{\langle \Gamma_1,\Gamma_2 \rangle}{2 \,
\Im(e^{-i \alpha} Z_1)_{r=\infty}}\,.
\end{equation}
Obviously, the separation has to be positive, so positivity of the
right hand side gives a necessary condition on the moduli at
spatial infinity for existence of a solution. It is indeed common
in $\CN=2$ theories for BPS states to exist only in certain
regions of moduli space. When one goes from a region where the
state exist to one where it doesn't, the BPS state decays into its
constituents, which is energetically only possible on a wall of
marginal stability, where the phases of the central charges of the
constituents align, that is $\arg Z_1 = \arg Z_2$. From
(\ref{eqsep}) it follows indeed that the separation diverges when
such a wall is approached.

\sect{Solutions}

\subsection{Solutions for $U$ and $z$: general case}

We now turn to the construction of explicit solutions. We start by
showing that if we use certain preferred holomorphic coordinates
on $\widetilde{\CM}$, solutions to (\ref{mc1}) can be expressed in
terms of a single function $\Sigma$ on $H^3(X,\IR)$, proportional to
the Bekenstein-Hawking entropy function. Consider first more
generally for an arbitrary $Q \in H^3(X,\IR)$ the equation
\begin{equation} \label{geneq}
 2 \, \Im (\bar{C} \Omega) = - Q
\end{equation}
in $n_v+1$ unknown complex variables, $z^a$ and $C$. Expressed in
components by computing intersection products with a basis
$\{\Theta_A\}$, this becomes
\begin{equation}
 2 \, \Im (\bar{C} Z_A) = - I_{AB} Q^B.
\end{equation}
These are $2n_v+2$ real equations, so we can in general expect a
finite number of solutions $C_*(Q)$, $z_*(Q)$ for a given $Q$. Now
note that by taking the intersection product of (\ref{geneq}) with
$\Omega$ and using $\langle \bar{\Omega},\Omega \rangle = i$, we
get $C_*(Q) = Z_Q |_{z_*(Q)} = Q^A Z_A|_{z_*(Q)}$, while taking
intersection products with $D_a \Omega$ and using $\langle
\Omega,D_a \Omega \rangle = \langle \bar{\Omega},D_a \Omega
\rangle = 0$ results in $Q^A \, D_a Z_A|_{z_*(Q)} = 0$, which in
turn implies $\partial_a |Q^A Z_A|^2 |_{z_*(Q)}=0$. Similarly
$\bpartial_{\bar{a}} |Q^A Z_A|^2 |_{z_*(Q)}=0$. Defining the
function\footnote{If the solution $C_*(Q)$, $z_*(Q)$ is not
unique, $\Sigma(Q)$ will be multi-valued. This will generically be the case
when $\CM$ has conifold-type singularities \cite{DGR}. Also, for a range values of
$Q$, there may be no solution at all.}
\begin{equation} \label{Fdef}
 \Sigma(Q) \equiv |Q^A Z_A|^2 |_{z_*(Q)} = |C_*(Q)|^2,
\end{equation}
we have
\begin{eqnarray*}
 \frac{\partial \Sigma}{\partial Q^A} &=&
 \left. \frac{\partial |Q^A Z_A|^2}{\partial Q^A} \right|_{z_*(Q)} +
 \partial_a |Q^A Z_A|^2 |_{z_*(Q)} \, \frac{\partial z_*^a}{\partial Q^A} +
 \bpartial_{\bar{a}} |Q^A Z_A|^2 |_{z_*(Q)} \, \frac{\partial \bz_*^{\bar{a}}}{\partial Q^A}
 \\
 &=& (\bar{Z}_Q Z_A + Z_Q \bar{Z}_A)|_{z_*(Q)} + 0 + 0 \\
 &=& 2 \, \Re(\bar{C}_* Z_A|_{z_*(Q)}).
\end{eqnarray*}
Therefore we find that a solution to (\ref{geneq}) for a given $Q$
satisfies
\begin{equation} \label{Zsol}
 2 \bar{C} Z_A = \partial_A \Sigma - i\,I_{AB} Q^B,
\end{equation}
and, since $Z_A = e^{\CK/2} X_A$:
\begin{equation} \label{modsol}
 t_A \equiv \frac{X_A}{X_0} =
 \frac{\partial_A \Sigma - i\, I_{AB} Q^B}{\partial_0 \Sigma - i\, I_{0B}
 Q^B},
\end{equation}
where the $0$-index refers to some suitably chosen basis element.
Locally, $n_v$ of the $t_A$ can be used as holomorphic coordinates
on $\widetilde{\CM}$. They are the usual ``special'' coordinates of
special geometry. Thus, (\ref{modsol}) together with (\ref{Fdef})
gives the values for $|C|$ and the moduli solving (\ref{geneq}) in
terms of a single function $\Sigma$ (which of course may still be hard
to compute). Note that $\Sigma$ is a homogeneous function of degree
two, i.e.\ $\Sigma(\lambda Q)=\lambda^2 \Sigma(Q)$, since (\ref{Zsol})
implies
\begin{equation}
 Q^A \partial_A \Sigma = 2 \bar{C} Q^A Z_A = 2 |Q^A Z_A|^2 = 2 \Sigma.
\end{equation}

Now we apply all this to solve the BPS field equation (\ref{mc1}),
which is of the form (\ref{geneq}) with $Q=H(\sx)$ and
$C=e^{-U(\sx)} e^{i \alpha(\sx)}$, so
\begin{equation} \label{mc1sol}
 e^{-2 U} = \Sigma(H) \quad \mbox{ and } \quad
 t_A \equiv \frac{X_A}{X_0} =
 \frac{\partial_A \Sigma(H) - i\, I_{AB} H^B}{\partial_0 \Sigma(H)- i\, I_{0B}
 H^B}.
\end{equation}
If $\Sigma$ is multi-valued, the relevant branch is selected by
continuity and the fact that the solution is unambiguous at
infinity (since it is given by the boundary conditions).
An obvious necessary
condition for existence of the solution is furthermore that $H$ stays within the
domain of $\Sigma$ everywhere.

To connect $\Sigma$ to the Bekenstein-Hawking entropy, consider the
case of a single charge $\Gamma$ at the origin, i.e. $H = \Gamma/r
- C$. Then we find for the horizon area of the BPS black hole thus
produced, from (\ref{mc1sol}) and the degree two homogeneity
property of $\Sigma$:
\begin{equation}
 A = 4 \pi \lim_{r \to 0} r^2 e^{-2U} = 4 \pi \lim_{r \to 0} r^2
 \Sigma(\Gamma/r-C) = 4 \pi \Sigma(\Gamma),
\end{equation}
and therefore for the Bekenstein-Hawking entropy function $S_{BH}$
\begin{equation}
 S_{BH}(\Gamma) = \pi \Sigma(\Gamma).
\end{equation}
So we see that if $S_{BH}$ is explicitly known, the full solution
(for $U$ and the moduli in special coordinates) can be constructed
explicitly as well. As a side remark, note that homogeneity of $\Sigma$
and (\ref{mc1sol}) similarly imply that the moduli at the horizon
$r=0$ are fixed to the value obtained by replacing $H \to \Gamma$
in the expression for $t_A$, and in particular are independent of
the moduli at infinity (apart for possible $\Sigma$-branch selection).
This is the well-known attractor mechanism of $\CN=2$ BPS black
holes \cite{FKS}.

Computing $S_{BH}$ in $\CN=2$ supergravity theories analytically
can be quite involved, and has only been done in limits where the
periods become polynomial in the special coordinates (e.g.\ large
complex structure (radius) limits in compactification of IIB (IIA)
on a Calabi-Yau) \cite{shmakova,M}. This usually proceeds through
solving an equation of the form (\ref{geneq}) (with $Q=\Gamma$),
and along the way one derives the corresponding expressions for
the $t_A$ as well. In practice it is often more convenient to
directly use those formulas rather than computing the $t_A$ from
$\Sigma$ using (\ref{mc1sol}). The only point of this section is then
the prescription that the full solution is obtained from the
horizon computations essentially by substituting the harmonic
function $H$ for the charge $\Gamma$.

Alternatively, in some cases, one can compute the entropy function
microscopically \cite{SV,vafaCYBH,MSW}. Then this section gives a
recipe to construct the full supergravity solution just from this
piece of microscopic information. In fact, using (\ref{modsol}),
one can in principle reconstruct the full special geometry from
knowledge of the entropy function alone.

\subsection{Solutions for $U$ and $z$: diagonal $T^6$ example}

In the case of the diagonal $T^6$ example of section
\ref{sec:diagT6}, the function $\Sigma$ is given by $\Sigma(Q)=\sqrt{D(Q)}$,
with $D(Q)$ the discriminant function \cite{M}:
\begin{equation} \label{T6dis}
 D = 3 p^2 q^2 + 4 p^3 u + 4 q^3 v + 6 u q p v - u^2 v^2
\end{equation}
where we have denoted the components of $Q$ with respect to the
$(D0,D2,D4,D6)$-basis as $(u,q,p,v)$, so $Q^A X_A = u + 3 q \tau -
3 p \tau^2 - v \tau^3$. For the corresponding modulus $\tau=b+a \,
i$ we get (using (\ref{modsol}) or directly from \cite{M}):
\begin{equation} \label{T6mod}
 \tau = b + a \, i = \frac{pq-uv + i \sqrt{D}}{2(p^2+q v)}.
\end{equation}
For charges $\Gamma_s = (\hat{u}_s,\hat{q}_s,\hat{p}_s,\hat{v}_s)$
at positions $\sx_s$, the harmonic function $H$ of (\ref{Haha})
is, with $r_s \equiv |\sx-\sx_s|$:
\begin{eqnarray}
 H(\sx) &=& \biggl(u(\sx),q(\sx),p(\sx),v(\sx)\biggr) \label{HT6a} \\
 &=&
 \biggl( \mbox{$\sum_s$} \frac{\hat{u}_s}{r_s} - c_0,
 \mbox{$\sum_s$} \frac{\hat{q}_s}{r_s} - c_2,
 \mbox{$\sum_s$} \frac{\hat{p}_s}{r_s} - c_4,
 \mbox{$\sum_s$} \frac{\hat{v}_s}{r_s} - c_6 \biggr), \label{HT6b}
\end{eqnarray}
where
\begin{equation}
 (c_0,c_2,c_4,c_6) = \frac{2}{\sqrt{8a^3}}
 \Im [e^{-i\alpha}(\tau^3,-\tau^2,-\tau,1)] \,\,\biggr|_{r=\infty}
\end{equation}
and $\alpha = \arg(\hat{u}+3 \hat{q} \tau -3 \hat{p} \tau^2 -
\hat{v} \tau^3 )$ with $(\hat{u},\hat{q},\hat{p},\hat{v}) = \sum_s
(\hat{u}_s,\hat{q}_s,\hat{p}_s,\hat{v}_s)$. The metric factor
$e^{-2U}$ of the solution is then obtained as $e^{-2U(\sx)} =
\sqrt{D(\sx)}$ with the $\sx$-dependent $(u,q,p,v)$ from
(\ref{HT6b}) plugged into (\ref{T6dis}), and the modulus field
$\tau(\sx)$ similarly from (\ref{T6mod}).

Note that for the solution to exist, we need $D \geq 0$
everywhere. In particular this requires $D(\Gamma_s) \geq 0$ for
all charges $\Gamma_s$. Recall furthermore that the constraint
(\ref{distconstr}) has to be satisfied.

A simple spherically symmetric example is provided by considering
a charge $(\hat{u},0,\hat{p},0)$, $\hat{u},\hat{p}>0$, with
$\tau_\infty = i a_\infty$. Then we get $e^{-2U}=\sqrt{D}=2
\sqrt{p^3u}$, $\tau=i \sqrt{u/p}$, with
$u=\hat{u}/r+\sqrt{a_\infty^3/2}$, $p=\hat{p}/r+1/\sqrt{2
a_\infty}$, reproducing the usual D0-D4 BPS black hole solution
with zero $B$-field.

\begin{figure}
  \epsfig{file=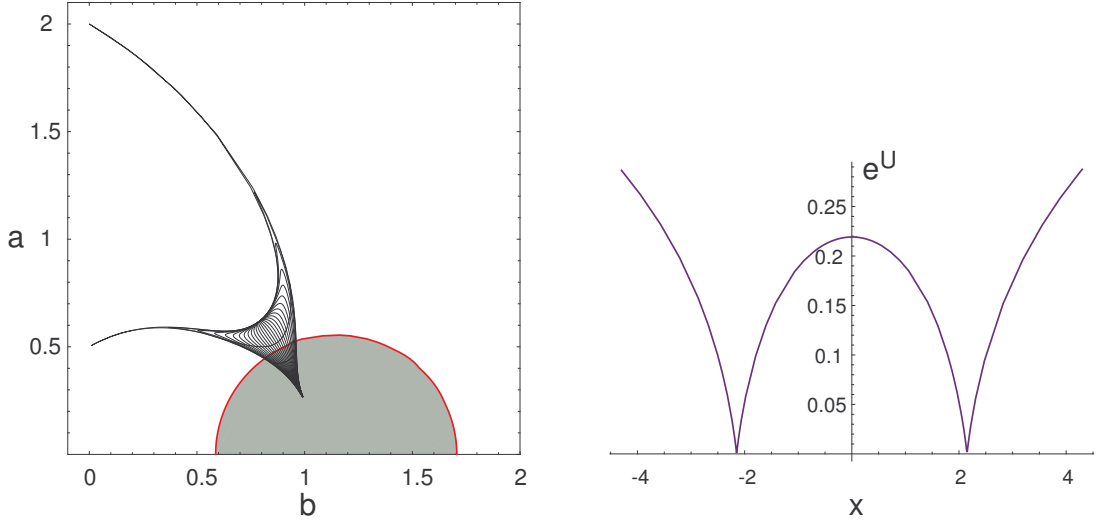,height=8cm,angle=0,trim=0 0 0 0}
  \caption{\emph{Left:} Profile of $\tau(\sx)$ in the upper half plane
  for the 2-centered example described in the text.
  The shaded region is the domain of allowed values of $\tau_\infty$.
  The red curve is the line of marginal stability. \emph{Right:} $e^U$ as a
  function of $x$ with the two charges located on the $x$-axis.}
  \label{tauplot}
\end{figure}

As a specific example of a two-centered solution, consider the
charges $\Gamma_1=(4,0,1,0)$ and $\Gamma_2=(0,4,0,1)$. Notice that
$D(\Gamma_1)=D(\Gamma_2)=16$ while $D(\Gamma_1+\Gamma_2)=-125$, so
a two-centered solution can indeed exist whereas a single centered
one cannot. From (\ref{eqsep}), we find for the separation of the
charges
\begin{equation}
 |\sx_1-\sx_2| = -19 \sqrt{2 a} |X| \, / \,
 (12a^4+12b^4+24a^2b^2+25a^2-39b^2+12) \biggr|_{r=\infty}
\end{equation}
with $X=X_1+X_2 = 4+12\tau-3\tau^2-\tau^3$. Therefore a necessary
condition for existence of the solution is
\begin{equation} \label{stabcond}
 12a^4+12b^4+24a^2b^2+25a^2-39b^2+12 < 0.
\end{equation}
The zero locus of the left hand side consists of two branches, one
with $b>0$ (shown in fig.\ \ref{tauplot}) and the other with
$b<0$. On the $b<0$ branch, we have $\arg Z_1 = \arg Z_2 + \pi$,
and on the $b>0$ branch, we have $\arg Z_1 = \arg Z_2$. As
discussed at the end of section \ref{sec:bpseom}, on general
grounds, only the latter branch can be a boundary between a region
where the BPS state exists and a region where it does not.
Therefore, the only region in which the BPS bound state can exist
is the region inside the branch with $b>0$.\footnote{When one
tries to construct a solution with $\tau$ at $r=\infty$ inside the
other branch, one finds that $D$ becomes negative and the solution
breaks down.}

To be completely explicit, let us pick the point $b=1$, $a=1/4$,
which lies inside the stable domain. The separation of the charges
is then $|\sx_1-\sx_2| \approx 4.29899$, and
\begin{equation}
 H \approx (\frac{4}{r_1}-5.47809,\frac{4}{r_2}+5.83671,
 \frac{1}{r_1}+5.83091,\frac{1}{r_2}-5.48245).
\end{equation}
The corresponding image of the map $\tau(\sx)$ is plotted in fig.\
\ref{tauplot}. It has the profile of a ``fattened split flow'', as
anticipated in \cite{branessugra,montreal}, with starting point
$\tau_\infty = 1+i/4$ and its two legs at $r_1=0$ and $r_2=0$
ending on the attractor points of respectively $\Gamma_1$ and
$\Gamma_2$, i.e.\ $\tau=2 i$ and $\tau=i/2$.

\subsection{Solutions for $\omega$}

We now turn to the solution of (\ref{mc2}). First we consider the
case of two centers, which generalizes straightforwardly to the
case of an arbitrary number of centers thanks to the linearity of
(\ref{mc2}). Let the position of the centers be $\sx_1=(0,0,-l)$
and $\sx_2=(0,0,l)$, and denote $\langle \Gamma_1,\Gamma_2 \rangle
\equiv \kappa$. Then according to the integrability condition
(\ref{eqsep}), we have $2 \Im(e^{-i \alpha} Z_1)_\infty = - 2
\Im(e^{-i \alpha} Z_2)_\infty = \kappa/2l$, so (\ref{mc2}) becomes
\begin{equation} \label{omeq2}
 d \omega = \kappa \s* \biggl( r_2^{-1} d\, r_1^{-1}
 - r_2^{-1} d\, r_1^{-1} - (2 l)^{-1} d\, r_1^{-1}
 + (2 l)^{-1} d\, r_2^{-1} \biggr).
\end{equation}
Introducing spherical coordinates $(r,\theta,\phi)$, and using the
identities
\begin{equation}
 r_1^2 = r^2+l^2+2 r l \cos \theta, \quad
 r_2^2 = r^2+l^2-2 r l \cos \theta, \quad
 (r_1 r_2)^2 = l^4 + r^4 - 2 l^2 r^2 \cos 2\theta
\end{equation}
and
\begin{equation}
 \s* dr = r^2 \sin \theta d\theta \wedge d\phi, \quad
 \s* d\theta = -\sin \theta dr \wedge d\phi,
\end{equation}
we get for (\ref{omeq2}):
\begin{eqnarray}
 d \omega &=& -\kappa \, l \, \frac{2 r(l^2+r^2)(1-\cos 2\theta) \, dr
 +r^2(l^2-r^2) \sin 2 \theta \, d\theta}{(l^4 + r^4 - 2l^2r^2\cos
 2\theta)^{3/2}}\wedge d\phi \nonumber \\
 &&+ \frac{\kappa}{2l} (\sin \theta_1 \, d\theta_1 - \sin \theta_2
 \, d \theta_2) \wedge d\phi  \nonumber \, ,
\end{eqnarray}
where $\theta_1$, $\theta_2$ are the angles with the $z$-axis in a
spherical coordinate system with origin at $\sx_1$ resp.\ $\sx_2$
(related to the central spherical coordinates by $r \cos \theta -
r_1 \cos \theta_1=l=r_2 \cos \theta_2 - r \cos \theta$).
Integrating this equation gives for $\omega$ (up to $\omega \to
\omega + d f$ gauge transformations):
\begin{equation} \label{omegasol2}
 \omega = \frac{\kappa}{2 l} \biggl(
 \frac{l^2-r^2}{(l^4+r^4-2l^2r^2\cos 2\theta)^{1/2}}
 + 1 - \cos \theta_1 + \cos \theta_2
 \biggr) d\phi \, .
\end{equation}
Note that the correct cancellations of terms occur on all segments
of the $z$-axis to make this solution non-singular. If we had not
implemented the integrability constraint (\ref{eqsep}), this would
not have been the case, and we would have had a physical
singularity on the $z$-axis.

Asymptotically for $r\to\infty$, we have $\omega \approx
\frac{\kappa}{r} \sin^2\theta \, d\phi = \frac{\kappa}{r^3}(x dy-y
dx)$, which implies \cite{MTW} that the angular momentum of this
spacetime is $\sJ=(0,0,\kappa/2)$, in agreement with the general
result of \cite{branessugra}.

In the case of an arbitrary number of centers $\sx_s$, satisfying
the constraint (\ref{distconstr}), we get the following for
(\ref{mc2}):
\begin{equation}
 \s* d\omega = \sum_{s<t} \kappa_{st} \biggl(
 r_t^{-1}d\,r_s^{-1} - r_s^{-1}d\,r_t^{-1}
 - R_{st}^{-1} d\,r_s^{-1} + R_{st}^{-1} d\,r_t^{-1}
 \biggr),
\end{equation}
where $R_{st}=|\sx_s-\sx_t|$. Clearly the solution to this
equation can be written as
\begin{equation}
 \omega = \sum_{s<t} \omega_{st}
\end{equation}
with $\omega_{st}$ the two-center solution of (\ref{omegasol2})
for the centers $\sx_s$ and $\sx_t$.

\subsection{Solution for $\CF$: general case}

Using (\ref{Zsol}) with $C=e^{i\alpha}e^{-U}$, and $e^{-2U}=\Sigma(H)$,
expression (\ref{bpsmc3}) for the total electromagnetic
field $\CF=d\CA$ can be written as follows:
\begin{equation} \label{Fssol}
 \CA^A = I^{AB} \partial_B \ln \Sigma(H) \, (dt + \omega) + \CA^A_d,
\end{equation}
where, introducing spherical coordinates $(r_s,\theta_s,\phi_s)$ around each center $\sx_s$,
\begin{equation}
 \CA_d = - \sum_s \cos \theta_s \, d \phi_s \otimes \Gamma_s.
\end{equation}
For a probe particle with charge $\Gamma_p$ in this background,
the second term in (\ref{Sp}) becomes
\begin{equation}
 \frac{1}{2} \int \langle \Gamma_p,\CA \rangle = \frac{1}{2} \int \Gamma_p^A \partial_A \ln \Sigma(H) \, (dt+\omega) +
 \langle \Gamma_p,\CA_d \rangle
\end{equation}
Note also that (\ref{Sp}) and (\ref{bpsmc3}) imply that the
potential for a static probe is
\begin{equation}
 V = 2 e^U |Z_p| \sin^2[(\alpha_p-\alpha)/2]
\end{equation}
where $\alpha_p = \arg Z_p$. The probe is therefore in (BPS)
equilibrium wherever $\alpha_p=\alpha$.

\sect{Dependence of entropy on moduli}

The existence of multi-centered black hole bound states implies
that entropy is not determined by charge only. If, say, a certain
charge supports a spherically symmetric black hole solution, and
in a certain region of moduli space also a multi-centered black
hole solution, the entropy associated to the charge will jump when
crossing into that region.

Here we give an explicit example of this phenomenon. Consider the
charges $Q_1 = (0,q,0,v)$, $Q_2=(u,0,0,0)$, $Q = Q_1 + Q_2$, with
$u,q,v>0$. The corresponding periods are $X_1=3 q \tau - v
\tau^3$, $X_2=u$, the discriminants are
\begin{eqnarray}
 D(Q_1) &=& 4 q^3 v \\
 D(Q_2) &=& 0 \\
 D(Q) &=& 4 q^3 v - u^2v^2,
\end{eqnarray}
and the attractor points
\begin{eqnarray}
 \tau_*(Q_1) &=& i \sqrt{\frac{q}{v}} \\
 \tau_*(Q_2) &=& 0 \\
 \tau_*(Q) &=& \frac{-uv+i\sqrt{4 q^3 v - u^2v^2}}{2qv} \, .
\end{eqnarray}
If $D(Q)>0$, a single-centered black hole solution always exists.
To find the region in the upper half plane where a two-centered
solution exists as well, we compute the $(Q_1,Q_2)$ line of
marginal stability $\Im(X_1 \bar{X}_2)=0$, $\Re(X_1 \bar{X}_2)>0$.
Writing $\tau=b+ia$:
\begin{eqnarray}
 \Im(X_1 \bar{X}_2) &=& u a (3 q - v(3b^2-a^2)) \\
 \Re(X_1 \bar{X}_2) &=& 8 u v b (b^2 - \frac{3}{4} \frac{q}{v}),
\end{eqnarray}
so the marginal stability line is the hyperbole branch
\begin{equation}
 a = \sqrt{3} \sqrt{b^2-\frac{q}{v}}, \quad b>\sqrt{\frac{q}{v}}
 \,.
\end{equation}
The condition for stability is
\begin{equation}
 \langle Q_1,Q_2 \rangle \, \Im(X_1 \bar{X}_2) > 0 \, .
\end{equation}
Because $\langle Q_1,Q_2 \rangle = - u v$, this implies that the
stable region is the region to the right of the MS line. In type
IIA language, this means we have to turn on a sufficiently large
B-field to have a stable two-centered BPS solution.\footnote{or
more generally a multicentered solution consisting of a core black
hole of charge $Q_1$ and a cloud of charge $(1,0,0,0)$ particles
(i.e.\ D0-branes) on a sphere with radius equal to the equilibrium
distance.} Note that $\sqrt{D(Q)}<\sqrt{D(Q_1)}+\sqrt{D(Q_2)}$, so
the two-centered solution with charges $Q_1$ and $Q_2$, if it
exists, has in fact more entropy than the single centered
solution.

The microscopic prediction which follows from this is that the
moduli space of this D-brane system will develop a new branch at a
certain critical value of the B-field (in IIA), with an
exponentially larger cohomology than the original branch.

\sect{Conclusions}

We have shown how non-static multi-centered BPS solutions of
$\CN=2$ supergravity can be constructed analytically. In
particular we argued that this can be done explicitly whenever the
BPS entropy as a function of charge is known explicitly. This
allowed us to verify directly some properties of these solutions
inferred earlier in \cite{branessugra,montreal}. We also gave an
explicit example of moduli-dependent entropy.

\vskip 10mm \noindent {\bf \large Acknowledgements}

\vskip 5mm \noindent

We would like to thank Bobby Acharya, Mike Douglas, Brian Greene
and Greg Moore for stimulating discussions.

\end{document}